\documentclass[12pt]{article}
\usepackage{amsmath,amssymb}
\usepackage{bm}
\usepackage{mathrsfs}
\usepackage{fourier}
\usepackage{multirow}
\allowdisplaybreaks[4]
\usepackage[usenames]{color}
\newcommand{{\Slashp}}{p\!\!\!\!\!\big/}
\newcommand{{\Slashq}}{q\!\!\!\!\!\big/}
\newcommand{{\Slashk}}{k\!\!\!\!\!\big/}
\newcommand{\LBox}{\mbox{\large$\Box$}}

\begin{document}

\title{Quantization of systems with $OSp(2|2)$ symmetry}

\author{
Yoshiharu \textsc{Kawamura}\footnote{E-mail: haru@azusa.shinshu-u.ac.jp}\\
{\it Department of Physics, Shinshu University, }\\
{\it Matsumoto 390-8621, Japan}\\
}

\date{
February 3, 2015}

\maketitle
\begin{abstract}
We study the quantization of systems that contain
both ordinary fields with a positive norm
and their counterparts obeying different statistics.
The systems have novel fermionic symmetries different from
the space-time supersymmetry and the BRST symmetry.
The unitarity of systems holds by imposing subsidiary conditions on states.
\end{abstract}


\section{Introduction}

The spin-statistics theorem explains that {\it observed particles of integer spin
obey Bose-Einstein statistics and are quantized by the commutation relations,
and those of half odd integer spin obey Fermi-Dirac statistics and are quantized
by the anti-commutation relations in the framework of 
relativistic quantum field theory}~\cite{P&B,deWet,P1,Feyn1,Feyn2,P2,Sch,L&Z,Bur,S&W,Ohta,D&S,F}.
The study on abnormal fields has been little carried out~\cite{F2,F3,Toth,YK},
except for Faddeev-Popov ghosts, i.e., ghost fields appearing on the quantization of
systems with local symmetries~\cite{F&P}.
Here, abnormal fields mean particles obeying different statistics from ordinary ones.
We refer to a scalar field following anti-commutation relations as a $\lq$fermionic scalar field'
and to a spinor field following commutation relations as a $\lq$bosonic spinor field'.

The reasons for the indifference of abnormal fields would be as follows.
First, they seem unrealistic because the standard model does not
contain abnormal ones irrelevant to gauge symmetries.
Second, in the introduction of abnormal fields, 
states with a negative norm appear
and the unitarity of systems can be violated.
Third, even if such unfavorable states are projected out by imposing subsidiary conditions
on states, abnormal fields become unphysical and cannot give any effects on physical processes.
Hence, we suppose that 
the existence of abnormal fields cannot be verified directly
or this is the same as the non-existence.

Nevertheless, it would be meaningful to examine systems 
with abnormal fields from following reasons.
There is a possibility that unphysical objects exist in nature 
if they are not prohibited from the consistency of theories.
This is a similar idea to that Dirac predicted 
the existence of magnetic monopole
based on quantum theory.
Unphysical ones might play a vital role
at a more fundamental level.
Furthermore, it is expected that they might leave some fingerprints
and we could check them as indirect proofs.

This paper takes a scholarly look at the nature of abnormal fields.
We study the quantization of systems 
that contain both ordinary fields with a positive norm
and their counterparts obeying different statistics.
We find that the systems have fermionic symmetries 
and the unitarity of systems holds by imposing
subsidiary conditions on states.
The fermionic symmetries are novel ones 
on a space of quantum fields, 
different from the space-time supersymmetry
and the BRST symmetry.

The content of this paper are as follows.
We study the quantization of system of scalar fields with $OSp(2|2)$ symmetry in Sect. II
and spinor fields with fermionic symmetries in Sect. III.
Section IV is devoted to conclusions and discussions.

\section{Systems of scalar fields with $OSp(2|2)$ symmetry}

Let us study the system that an ordinary complex scalar field $\varphi$
and the fermionic one $c_{\varphi}$ coexist, described by the Lagrangian density,
\begin{eqnarray}
\mathcal{L}_{\varphi, c_{\varphi}} 
= \partial_{\mu} \varphi^{\dagger} \partial^{\mu} \varphi - m^2 \varphi^{\dagger} \varphi
+ \partial_{\mu} c_{\varphi}^{\dagger} \partial^{\mu} c_{\varphi} - m^2 c_{\varphi}^{\dagger} c_{\varphi}.
\label{L-varphi-c}
\end{eqnarray}
Based on the formulation with the property that
{\it the hermitian conjugate of canonical momentum for a variable is just the canonical momentum
for the hermitian conjugate of the variable}, 
we define the conjugate momentum of $\varphi$, $\varphi^{\dagger}$, 
$c_{\varphi}$ and $c_{\varphi}^{\dagger}$ as
\begin{eqnarray}
&~& \pi \equiv \left(\frac{\partial \mathcal{L}_{\varphi, c_{\varphi}}}{\partial \dot{\varphi}}\right)_{\rm R}
=  \dot{\varphi}^{\dagger},~~
\pi^{\dagger} \equiv 
\left(\frac{\partial \mathcal{L}_{\varphi, c_{\varphi}}}{\partial \dot{\varphi}^{\dagger}}\right)_{\rm L}
= \dot{\varphi},
\label{pi}\\
&~& \pi_{c_{\varphi}} \equiv 
\left(\frac{\partial \mathcal{L}_{\varphi, c_{\varphi}}}{\partial \dot{c}_{\varphi}}\right)_{\rm R}
=  \dot{c}_{\varphi}^{\dagger},~~
\pi_{c_{\varphi}}^{\dagger} \equiv 
\left(\frac{\partial \mathcal{L}_{\varphi, c_{\varphi}}}{\partial \dot{c}_{\varphi}^{\dagger}}\right)_{\rm L}
= \dot{c}_{\varphi},
\label{pi-c}
\end{eqnarray}
where R and L stand for the right-differentiation and the left-differentiation, respectively.

By solving the Klein-Gordon equations $\left(\raisebox{-0.6mm}{\LBox} + m^2\right) \varphi = 0$
and $\left(\raisebox{-0.6mm}{\LBox} + m^2\right) c_{\varphi} = 0$, we obtain the solutions
\begin{eqnarray}
&~& \varphi(x) = \int \frac{d^3k}{\sqrt{(2\pi)^3 2k_0}}
\left(a(\bm{k}) e^{-i k x} + b^{\dagger}(\bm{k}) e^{i k x}\right),
\label{varphi-sol}\\
&~& \varphi^{\dagger}(x) = \int \frac{d^3k}{\sqrt{(2\pi)^3 2k_0}}
\left(a^{\dagger}(\bm{k}) e^{i k x} + b (\bm{k}) e^{-i k x}\right),
\label{varphi-dagger-sol}\\
&~& \pi(x) = i \int d^3k \sqrt{\frac{k_0}{2 (2\pi)^3}}
\left(a^{\dagger}(\bm{k}) e^{i k x} - b (\bm{k}) e^{-i k x}\right),
\label{pi-sol}\\
&~& \pi^{\dagger}(x) = - i \int d^3k \sqrt{\frac{k_0}{2 (2\pi)^3}}
\left(a(\bm{k}) e^{-i k x} - b^{\dagger} (\bm{k}) e^{i k x}\right),
\label{pi-dagger-sol}\\
&~& c_{\varphi}(x) = \int \frac{d^3k}{\sqrt{(2\pi)^3 2k_0}}
\left(c(\bm{k}) e^{-i k x} + d^{\dagger}(\bm{k}) e^{i k x}\right),
\label{c-sol}\\
&~& c_{\varphi}^{\dagger}(x) = \int \frac{d^3k}{\sqrt{(2\pi)^3 2k_0}}
\left(c^{\dagger}(\bm{k}) e^{i k x} + d (\bm{k}) e^{-i k x}\right),
\label{c-dagger-sol}\\
&~& \pi_{c_{\varphi}}(x) = i \int d^3k \sqrt{\frac{k_0}{2 (2\pi)^3}}
\left(c^{\dagger}(\bm{k}) e^{i k x} - d (\bm{k}) e^{-i k x}\right),
\label{pi-c-sol}\\
&~& \pi_{c_{\varphi}}^{\dagger}(x) = - i \int d^3k \sqrt{\frac{k_0}{2 (2\pi)^3}}
\left(c(\bm{k}) e^{-i k x} - d^{\dagger} (\bm{k}) e^{i k x}\right),
\label{pi-c-dagger-sol}
\end{eqnarray}
where $k_0 = \sqrt{\bm{k}^2 + m^2}$ and $kx = k^{\mu} x_{\mu}$.

Using (\ref{pi}) and (\ref{pi-c}), the Hamiltonian density is obtained as
\begin{eqnarray}
&~& \mathcal{H}_{\varphi, c_{\varphi}} = \pi \dot{\varphi} 
+ \dot{\varphi}^{\dagger}\pi^{\dagger} + \pi_{c_{\varphi}} \dot{c}_{\varphi} 
+ \dot{c}_{\varphi}^{\dagger} \pi_{c_{\varphi}}^{\dagger}
- \mathcal{L}_{\varphi, c_{\varphi}}
\nonumber \\
&~& ~~~~~~~~~~~~ = \pi \pi^{\dagger} 
+ \bm{\nabla} \varphi^{\dagger} \bm{\nabla} \varphi
+ m^2 \varphi^{\dagger} \varphi
+ \pi_{c_{\varphi}} \pi_{c_{\varphi}}^{\dagger} 
+ \bm{\nabla} c_{\varphi}^{\dagger} \bm{\nabla} c_{\varphi}
+ m^2 c_{\varphi}^{\dagger} c_{\varphi}.
\label{H-varphi-c}
\end{eqnarray}

The system is quantized by regarding variables as operators
and imposing the following relations 
on the canonical pairs $(\varphi, \pi)$, $(\varphi^{\dagger}, \pi^{\dagger})$,
$(c_{\varphi}, \pi_{c_{\varphi}})$ 
and $(c_{\varphi}^{\dagger}, \pi_{c_{\varphi}}^{\dagger})$,
\begin{eqnarray}
&~& [\varphi(\bm{x}, t), \pi(\bm{y}, t)] = i \delta^3(\bm{x}-\bm{y}),~~ 
[\varphi^{\dagger}(\bm{x}, t), \pi^{\dagger}(\bm{y}, t)] = i \delta^3(\bm{x}-\bm{y}),
\label{CCR-varphi}\\
&~& \{c_{\varphi}(\bm{x}, t), \pi_{c_{\varphi}}(\bm{y}, t)\} = i \delta^3(\bm{x}-\bm{y}),~~ 
\{c_{\varphi}^{\dagger}(\bm{x}, t), \pi_{c_{\varphi}}^{\dagger}(\bm{y}, t)\} = -i \delta^3(\bm{x}-\bm{y}),
\label{CCR-c}
\end{eqnarray}
where $[O_1, O_2] \equiv O_1 O_2 - O_2 O_1$,
$\{O_1, O_2\} \equiv O_1 O_2 + O_2 O_1$,
and only the non-vanishing ones are denoted.
Or equivalently, the following relations are imposed on,
\begin{eqnarray}
&~& [a(\bm{k}), a^{\dagger}(\bm{l})] = \delta^3(\bm{k}-\bm{l}),~~ 
[b(\bm{k}), b^{\dagger}(\bm{l})] = \delta^3(\bm{k}-\bm{l}), 
\label{CCR-ab-varphi}\\
&~& \{c(\bm{k}), c^{\dagger}(\bm{l})\} = \delta^3(\bm{k}-\bm{l}),~~
\{d(\bm{k}), d^{\dagger}(\bm{l})\} = - \delta^3(\bm{k}-\bm{l}), 
\label{CCR-cd-c}
\end{eqnarray}
and others are zero.

By inserting (\ref{varphi-sol}) -- (\ref{pi-c-dagger-sol}) into (\ref{H-varphi-c}),
the Hamiltonian ${H}_{\varphi, c_{\varphi}}$ is written by
\begin{eqnarray}
H_{\varphi, c_{\varphi}} = \int \mathcal{H}_{\varphi, c_{\varphi}} d^3x
= \int d^3k k_0 
\left(a^{\dagger}(\bm{k}) a(\bm{k}) + b^{\dagger}(\bm{k}) b(\bm{k})
+ c^{\dagger}(\bm{k}) c(\bm{k}) - d^{\dagger}(\bm{k}) d(\bm{k})\right).
\label{H-abcd-varphi-c}
\end{eqnarray} 
Note that the sum of the zero-point energies vanishes due to the cancellation between 
contributions from  $(\varphi, \varphi^{\dagger})$ and $(c_{\varphi}, c_{\varphi}^{\dagger})$.

The eigenstates for $H_{\varphi, c_{\varphi}}$ are constructed by acting
the creation operators $a^{\dagger}(\bm{k})$, $b^{\dagger}(\bm{k})$,
$c^{\dagger}(\bm{k})$ and $d^{\dagger}(\bm{k})$ on the vacuum state $| 0 \rangle$,
where $| 0 \rangle$ is defined by the conditions $a(\bm{k})| 0 \rangle = 0$,
$b(\bm{k})| 0 \rangle = 0$, $c(\bm{k})| 0 \rangle = 0$ and $d(\bm{k})| 0 \rangle = 0$.
We find that the energy is positive semi-definite,
because the effect on the negative sign appearing in front of $d^{\dagger}(\bm{k}) d(\bm{k})$ in ${H}_{\varphi, c_{\varphi}}$
changes into an opposite one by the negative sign in the relation 
$\{d(\bm{k}), d^{\dagger}(\bm{l})\} = -\delta^3(\bm{k}-\bm{l})$.

The microscopic causality also holds seen from the 4-dimensional relations as
\begin{eqnarray}
&~& [\varphi(x), \varphi^{\dagger}(y)] 
= \{c_{\varphi}(x), c_{\varphi}^{\dagger}(y)\}
= \int \frac{d^3k}{(2\pi)^3 2k_0} \left(e^{-ik(x-y)} - e^{ik(x-y)}\right)
\nonumber \\
&~& ~~~~~~~~~~~~~~~~~~~~~~~~ 
= \int \frac{d^4k}{(2\pi)^3} \epsilon(k_0) \delta(k^2-m^2) e^{-ik(x-y)} 
\equiv i \varDelta(x-y),~~ 
\label{4D-CCR1}\\
&~& [\varphi(x), \varphi(y)] = 0,~~
[\varphi^{\dagger}(x), \varphi^{\dagger}(y)] = 0,~~
\{c_{\varphi}(x), c_{\varphi}(y)\} = 0,~~
\{c_{\varphi}^{\dagger}(x), c_{\varphi}^{\dagger}(y)\} = 0,
\label{4D-CCR2}\\
&~& [\varphi(x), c_{\varphi}(y)] = 0,~~
[\varphi(x), c_{\varphi}^{\dagger}(y)] = 0,~~
[\varphi^{\dagger}(x), c_{\varphi}(y)] = 0,~~
[\varphi^{\dagger}(x), c_{\varphi}^{\dagger}(y)] = 0,
\label{4D-CCR3}
\end{eqnarray}
where $\epsilon(k_0) = k_0/|k_0|$ with $\epsilon(0)=0$,
$\varDelta(x-y)$ is the invariant delta function, and
two fields separated by a space-like interval commute or anti-commute with each other
as seen from the relation $\varDelta(x-y) = 0$ for $(x-y)^2 < 0$.
Note that bosonic variables composed of $c_{\varphi}$ and $c_{\varphi}^{\dagger}$
are commutative to any bosonic variables separated by a space-like interval.

The system contains negative norm states originated from
$\{d(\bm{k}), d^{\dagger}(\bm{l})\} = -\delta^3(\bm{k}-\bm{l})$.
For instance, from the relation,
\begin{eqnarray}
&~& 0 < \int d^3k \left|f(\bm{k})\right|^2 
= - \int d^3k \int d^3l f(\bm{k})^{*} f(\bm{l}) 
\langle 0 |\{d(\bm{k}), d^{\dagger}(\bm{l})\}| 0 \rangle
\nonumber \\
&~& ~~~~~~~~~ = - \int d^3k \int d^3l f(\bm{k})^{*} f(\bm{l}) 
\langle 0 |d(\bm{k})d^{\dagger}(\bm{l})| 0 \rangle
= - \left|\int d^3k f(\bm{k})d^{\dagger}(\bm{k})| 0 \rangle\right|^2,
\label{f-square}
\end{eqnarray} 
we see that the state $\int d^3k f(\bm{k})d^{\dagger}(\bm{k})| 0 \rangle$ has a negative norm.
Here, $f(\bm{k})$ is some square integrable functions.
In the presence of negative norm states,
the probability interpretation cannot be endured.
In the following, it is shown that the system has fermionic symmetries
and they can guarantee the unitarity of the system.

Now, let us investigate the symmetries of the system.
The $\mathcal{L}_{\varphi, c_{\varphi}}$ is invariant under the transformations
whose generators are the Lie algebras of $OSp(2|2)$.
In the appendix A, we explain more about $OSp(2|2)$ and $OSp(1,1|2)$
and field theories with such symmetries.

The transformations form following types.\\
(a) $U(1)$ transformation relating $\varphi$ and $\varphi^{\dagger}$:
\begin{eqnarray}
\delta_{\rm o} \varphi = - iq \epsilon_{\rm o} \varphi,~~
\delta_{\rm o} \varphi^{\dagger} = iq \epsilon_{\rm o} \varphi^{\dagger},~~
\delta_{\rm o} c_{\varphi} = 0,~~
\delta_{\rm o} c_{\varphi}^{\dagger} = 0,
\label{U(1)-varphi}
\end{eqnarray}
where $q$ is a $U(1)$ charge of $\varphi$ and $\epsilon_{\rm o}$ is an infinitesimal real number.\\
(b) $U(1)$ transformation relating $c_{\varphi}$ and $c_{\varphi}^{\dagger}$:
\begin{eqnarray}
\delta_{\rm g} \varphi = 0,~~
\delta_{\rm g} \varphi^{\dagger} = 0,~~
\delta_{\rm g} c_{\varphi} = -iq \epsilon_{\rm g} c_{\varphi},~~
\delta_{\rm g} c_{\varphi}^{\dagger} = iq \epsilon_{\rm g} c_{\varphi}^{\dagger},
\label{U(1)-c-varphi}
\end{eqnarray}
where $q$ is a $U(1)$ charge of $c_{\varphi}$ and $\epsilon_{\rm g}$ is an infinitesimal real number.\\
(c)Fermionic transformations:
\begin{eqnarray}
&~& \delta_{\rm F} \varphi = - r\zeta c_{\varphi},~~\delta_{\rm F} \varphi^{\dagger} = 0,~~ 
\delta_{\rm F} c_{\varphi} = 0,~~\delta_{\rm F} c_{\varphi}^{\dagger} = r \zeta \varphi^{\dagger},~~
\label{delta-F-varphi}\\
&~& \delta_{\rm F}^{\dagger} \varphi = 0,~~
\delta_{\rm F}^{\dagger} \varphi^{\dagger} = r \zeta^{\dagger} c_{\varphi}^{\dagger},~~
\delta_{\rm F}^{\dagger} c_{\varphi} = r \zeta^{\dagger} \varphi,~~
\delta_{\rm F}^{\dagger} c_{\varphi}^{\dagger} = 0,
\label{delta-Fdagger-varphi}
\end{eqnarray}
where $r = q^{1/2}$ and $\zeta$ and $\zeta^{\dagger}$ 
are Grassmann numbers.
Note that $\delta_{\rm F}$ and $\delta_{\rm F}^{\dagger}$
are not generated by hermitian operators, 
different from the generator of the BRST transformation
in systems with first class constraints~\cite{BRST} 
and that of the topological symmetry~\cite{W,Top}.

From the above transformation properties, 
we see that ${\bm \delta}_{\rm F}$ and ${\bm \delta}_{\rm F}^{\dagger}$
are nilpotent,
i.e., ${\bm{\delta}_{\rm F}}^2 = 0$ and ${\bm{\delta}_{\rm F}^{\dagger}}^2 = 0$
where ${\bm \delta}_{\rm F}$ and ${\bm \delta}_{\rm F}^{\dagger}$,
are defined by
$\delta_{\rm F} = \zeta {\bm \delta}_{\rm F}$
and
$\delta_{\rm F}^{\dagger} = \zeta^{\dagger} {\bm \delta}_{\rm F}^{\dagger}$,
respectively.
Furthermore, the following algebraic relations hold:
\begin{eqnarray}
{Q_{\rm F}}^2 = 0,~~{Q_{\rm F}^{\dagger}}^2 = 0,~~
\{Q_{\rm F}, Q_{\rm F}^{\dagger}\} = Q_{\rm o}+Q_{\rm g} \equiv N_{\rm D},
\label{QQdagger-varphi}
\end{eqnarray}
where $Q_{\rm F}$, $Q_{\rm F}^{\dagger}$, $Q_{\rm o}$ and $Q_{\rm g}$
are corresponding generators (charges) given by
\begin{eqnarray}
\delta_{\rm F} \Phi = i[\zeta Q_{\rm F}, \Phi],~~
\delta_{\rm F}^{\dagger} \Phi = i[Q_{\rm F}^{\dagger}\zeta^{\dagger}, \Phi],~~
\delta_{\rm o} \Phi = i[\epsilon_{\rm o} Q_{\rm o}, \Phi],~~
\delta_{\rm g} \Phi = i[\epsilon_{\rm g} Q_{\rm g}, \Phi].
\label{delta}
\end{eqnarray}

From the definition,
\begin{eqnarray}
&~& \zeta Q_{\rm F} 
\equiv \int d^3x \left[
\left(\frac{\partial \mathcal{L}_{\varphi, c_{\varphi}}}{\partial \dot{\varphi}}\right)_{\rm R} 
\delta_{\rm F} \varphi
+ \delta_{\rm F} c_{\varphi}^{\dagger} 
\left(\frac{\partial \mathcal{L}_{\varphi, c_{\varphi}}}{\partial \dot{c}_{\varphi}^{\dagger}}\right)_{\rm L}
\right],
\label{Q-F-def-varphi}\\
&~& Q_{\rm F}^{\dagger} \zeta^{\dagger} 
\equiv \int d^3x \left[
\delta_{\rm F}^{\dagger} \varphi^{\dagger}
\left(\frac{\partial \mathcal{L}_{\varphi, c_{\varphi}}}{\partial \dot{\varphi}^{\dagger}}\right)_{\rm L} 
+ \left(\frac{\partial \mathcal{L}_{\varphi, c_{\varphi}}}{\partial \dot{c}_{\varphi}}\right)_{\rm R}
\delta_{\rm F}^{\dagger} c_{\varphi}
\right],
\label{Q-F-dagger-def-varphi}
\end{eqnarray}
the conserved fermionic charges $Q_{\rm F}$ and $Q_{\rm F}^{\dagger}$ are obtained by
\begin{eqnarray}
&~& Q_{\rm F} = \int d^3x~r \left( - \pi c_{\varphi} + \varphi^{\dagger} \pi_{c_{\varphi}}^{\dagger}\right)
= - i \int d^3k~r \left(a^{\dagger}(\bm{k}) c(\bm{k}) - d^{\dagger}(\bm{k}) b(\bm{k})\right),~~
\label{QF-varphi}\\
&~& Q_{\rm F}^{\dagger} = \int d^3x~r \left(- c_{\varphi}^{\dagger} \pi^{\dagger}
+ \pi_{c_{\varphi}} \varphi\right)
= i \int d^3k~r \left(c^{\dagger}(\bm{k}) a(\bm{k}) - b^{\dagger}(\bm{k}) d(\bm{k})\right).
\label{QF-dagger-varphi}
\end{eqnarray}
Then, under the fermionic transformations, the canonical momenta are transformed as,
\begin{eqnarray}
&~& \delta_{\rm F} \pi = 0,~~\delta_{\rm F} \pi^{\dagger} = - r \zeta \pi_{c_{\varphi}}^{\dagger},~~ 
\delta_{\rm F} \pi_{c_{\varphi}}  = r \zeta \pi,~~
\delta_{\rm F} \pi_{c_{\varphi}}^{\dagger} = 0,
\label{delta-F-pi}\\
&~& \delta_{\rm F}^{\dagger} \pi = r \zeta^{\dagger} \pi_{c_{\varphi}},~~
\delta_{\rm F}^{\dagger} \pi^{\dagger} = 0,~~
\delta_{\rm F}^{\dagger} \pi_{c_{\varphi}} = 0,~~
\delta_{\rm F}^{\dagger} \pi_{c_{\varphi}}^{\dagger} = - r \zeta^{\dagger} \pi^{\dagger}.
\label{delta-Fdagger-pi}
\end{eqnarray}

The conserved $U(1)$ charge $N_{\rm D}$ is given by
\begin{eqnarray}
N_{\rm D} = \int d^3k~q \left(a^{\dagger}(\bm{k}) a(\bm{k}) - b^{\dagger}(\bm{k}) b(\bm{k})
+ c^{\dagger}(\bm{k}) c(\bm{k}) + d^{\dagger}(\bm{k}) d(\bm{k})\right).
\label{ND-varphi}
\end{eqnarray}
We find that the $U(1)$ charge of particle corresponding 
$b^{\dagger}(\bm{k})| 0 \rangle$ and $d^{\dagger}(\bm{k})| 0 \rangle$ is opposite to that corresponding 
$a^{\dagger}(\bm{k})| 0 \rangle$ and $c^{\dagger}(\bm{k})| 0 \rangle$.
Hence, $a(\bm{k})$ ($c(\bm{k})$) and $b^{\dagger}(\bm{k})$ ($d^{\dagger}(\bm{k})$) are regarded as 
the annihilation operator of particle (fermionic one) and the creation operator of antiparticle 
(antiparticle of fermionic one), respectively.

It is easily understood that $\mathcal{L}_{\varphi, c_{\varphi}}$ is invariant 
under the transformations (\ref{delta-F-varphi}) and (\ref{delta-Fdagger-varphi}),
from the nilpotency of ${\bm \delta}_{\rm F}$ and ${\bm \delta}_{\rm F}^{\dagger}$ 
and the relations,
\begin{eqnarray}
\mathcal{L}_{\varphi, c_{\varphi}} 
= {\bm \delta}_{\rm F} {\bm \delta}_{\rm F}^{\dagger} \left(\mathcal{L}_{\varphi}/q\right)
= - {\bm \delta}_{\rm F}^{\dagger} {\bm \delta}_{\rm F} \left(\mathcal{L}_{\varphi}/q\right),
\label{delta-rel-varphi}
\end{eqnarray}
where $\mathcal{L}_{\varphi}$ is given by
\begin{eqnarray}
\mathcal{L}_{\varphi} =  \partial_{\mu} \varphi^{\dagger} \partial^{\mu} \varphi - m^2 \varphi^{\dagger} \varphi.
\label{L-varphi}
\end{eqnarray}
The Hamiltonian density $\mathcal{H}_{\varphi, c_{\varphi}}$ is written in the $Q_{\rm F}$ and $Q_{\rm F}^{\dagger}$
exact forms such that 
\begin{eqnarray}
\mathcal{H}_{\varphi, c_{\varphi}} 
= \left\{Q_{\rm F}, \left\{Q_{\rm F}^{\dagger}, 
\mathcal{H}_{\varphi}/q\right\}\right\}
 = - \left\{Q_{\rm F}^{\dagger}, \left\{Q_{\rm F}, 
\mathcal{H}_{\varphi}/q\right\}\right\},
\label{H-varphi-c-exact}
\end{eqnarray}
where $\mathcal{H}_{\varphi}$ is given by
\begin{eqnarray}
\mathcal{H}_{\varphi} =  \pi \pi^{\dagger} 
+ \bm{\nabla} \varphi^{\dagger} \bm{\nabla} \varphi + m^2  \varphi^{\dagger} \varphi.
\label{H-varphi}
\end{eqnarray}

To formulate our model in a consistent manner,
we use a feature that {\it a conserved charge can be, in general, 
set to be zero as a subsidiary condition}.
We impose the following subsidiary conditions on states to select physical states,
\begin{eqnarray}
Q_{\rm F} |{\rm phys}\rangle = 0,~~
Q_{\rm F}^{\dagger} |{\rm phys}\rangle = 0,~~
N_{\rm D} |{\rm phys}\rangle = 0.
\label{Phys}
\end{eqnarray}
Note that $Q_{\rm F}^{\dagger} |{\rm phys}\rangle = 0$ means $\langle {\rm phys}|Q_{\rm F}=0$.
The conditions (\ref{Phys}) are interpreted 
as counterparts of the Kugo-Ojima subsidiary condition
in the BRST quantization~\cite{K&O1,K&O2}.
We find that all states, 
except for the vacuum state $|0 \rangle$, are
unphysical because they do not satisfy (\ref{Phys}).
This feature is understood as the quartet mechanism~\cite{K&O1,K&O2}.
The projection operator $P^{(n)}$ on the states with $n$ particles
is given by
\begin{eqnarray}
P^{(n)} = \frac{1}{n} \left(a^{\dagger} P^{(n-1)} a + b^{\dagger} P^{(n-1)} b + c^{\dagger} P^{(n-1)} c 
- d^{\dagger} P^{(n-1)} d \right)~~~~(n \ge 1),
\label{P(n)}
\end{eqnarray}
and is written by
\begin{eqnarray}
P^{(n)} =  i \left\{Q_{\rm F}, R^{(n)}\right\}~,
\label{P(n)2}
\end{eqnarray}
where $R^{(n)}$ is given by
\begin{eqnarray}
R^{(n)} = \frac{1}{n} \left(c^{\dagger} P^{(n-1)} a + b^{\dagger} P^{(n-1)} d\right)~~~~(n \ge 1).
\label{R(n)}
\end{eqnarray}
We find that any state with $n \ge 1$ is unphysical from the relation
$\langle {\rm phys}|P^{(n)}|{\rm phys}\rangle = 0$ for  $n \ge 1$.
Then, we understand that both $\varphi$ and $c_{\varphi}$ become unphysical,
and only $|0 \rangle$ is the physical one.
This is also regarded as a field theoretical version of the Parisi-Sourlas mechanism~\cite{P&S}.

The system is also described by hermitian fermionic charges
defined by $Q_1 \equiv Q_{\rm F} + Q_{\rm F}^{\dagger}$ and 
$Q_2 \equiv i(Q_{\rm F} - Q_{\rm F}^{\dagger})$.
They satisfy the relations $Q_1 Q_2 + Q_2 Q_1 = 0$,
${Q_1}^2 = N_{\rm D}$ and ${Q_2}^2 = N_{\rm D}$.
Though $Q_1$, $Q_2$ and $N_{\rm D}$
form elements of the $N=2$ (quantum mechanical) 
supersymmetry algebra~\cite{Witten},
our system does not possess the space-time supersymmetry
because $N_{\rm D}$ is not our Hamiltonian $H_{\varphi, c_{\varphi}}$
but the $U(1)$ charge $N_{\rm D}$.
Only the vacuum state is selected as the physical states
by imposing the following subsidiary conditions on states, 
in place of (\ref{Phys}),
\begin{eqnarray}
Q_{1} |{\rm phys}\rangle = 0,~~
Q_{2} |{\rm phys}\rangle = 0,~~
N_{\rm D} |{\rm phys}\rangle = 0.
\label{Phys-2}
\end{eqnarray}
It is also understood that our fermionic symmetries are different from 
the space-time supersymmetry,
from the fact that $Q_1$ and $Q_2$ are scalar charges.
They are also different from the BRST symmetry,
as seen from the algebraic relations among charges.

We discuss interactions among fields forming $Q_{\rm F}$-doublets.
Let us consider a system with two sets of $Q_{\rm F}$-doublet scalar fields 
($\varphi_1$, $c_{\varphi_1}$) and ($\varphi_2$, $c_{\varphi_2}$),
described by the Lagrangian density,
\begin{eqnarray}
&~& \mathcal{L}_{\varphi_i, c_{\varphi_i}} 
= \partial_{\mu} \varphi_1^{\dagger} \partial^{\mu} \varphi_1 - m_1^2 \varphi_1^{\dagger} \varphi_1
+ \partial_{\mu} c_{\varphi_1}^{\dagger} \partial^{\mu} c_{\varphi_1} - m_1^2 c_{\varphi_1}^{\dagger} c_{\varphi_1}
\nonumber \\
&~& ~~~~~~~~~~~~~~~~~ 
+ \partial_{\mu} \varphi_2^{\dagger} \partial^{\mu} \varphi_2 - m_2^2 \varphi_2^{\dagger} \varphi_2
+ \partial_{\mu} c_{\varphi_2}^{\dagger} \partial^{\mu} c_{\varphi_2} - m_2^2 c_{\varphi_2}^{\dagger} c_{\varphi_2}
\nonumber \\
&~& ~~~~~~~~~~~~~~~~~ 
- \lambda \left(\varphi_1^{\dagger} \varphi_1 
+ c_{\varphi_1}^{\dagger} c_{\varphi_1}\right)
\left(\varphi_2^{\dagger} \varphi_2 
+ c_{\varphi_2}^{\dagger} c_{\varphi_2}\right)
\nonumber \\
&~& ~~~~~~~~~~~~~~ 
= \bm{\delta}_{\rm F} \bm{\delta}_{\rm F}^{\dagger} 
\left(\partial_{\mu} \varphi_1^{\dagger} \partial^{\mu} \varphi_1 - m_1^2 \varphi_1^{\dagger} \varphi_1
+ \partial_{\mu} \varphi_2^{\dagger} \partial^{\mu} \varphi_2 - m_2^2 \varphi_2^{\dagger} \varphi_2
- \lambda \varphi_1^{\dagger} \varphi_1 
\varphi_2^{\dagger} \varphi_2\right),
\label{L-varphi-c-int}
\end{eqnarray}
where we take $q=1$ for simplicity.
We find that $\mathcal{L}_{\varphi_i, c_{\varphi_i}}$ 
does not receive any radiative corrections,
due to the cancellation between contributions 
from $\varphi_i$ and $c_{\varphi_i}$,
in the presence of interactions.
Or $Q_{\rm F}$-doublets interact with each other 
respecting the $OSp(2|2)$ invariance at the quantum level.
This system is also unrealistic, 
because all fields become unphysical 
and only the vacuum state survives as a physical one
after imposing subsidiary conditions on states.

\section{Systems of spinor fields with fermionic symmetries}

We study the system that an ordinary spinor field $\psi$
and its bosonic counterpart $c_{\psi}$ coexist, described by the Lagrangian density,
\begin{eqnarray}
\mathcal{L}_{\psi, c_{\psi}} 
= i \overline{\psi} \gamma^{\mu} \partial_{\mu} \psi - m \overline{\psi} \psi
+ i \overline{c}_{\psi} \gamma^{\mu} \partial_{\mu} c_{\psi} - m \overline{c}_{\psi}c_{\psi},
\label{L-psi-c}
\end{eqnarray}
where $\overline{\psi} \equiv \psi^{\dagger} \gamma^0$,
$\overline{c}_{\psi} \equiv c_{\psi}^{\dagger} \gamma^0$
and $\gamma^{\mu}$ are the gamma matrices satisfying 
$\{\gamma^{\mu}, \gamma^{\nu}\} = 2 \eta^{\mu\nu}$.

The canonical conjugate momentum of $\psi$ and $c_{\psi}$ are given by
\begin{eqnarray}
\pi_{\psi} \equiv \left(\frac{\partial \mathcal{L}_{\psi, c_{\psi}}}{\partial \dot{\psi}}\right)_{\rm R}
= i \overline{\psi} \gamma^0 = i \psi^{\dagger},~~ 
\pi_{c_{\psi}} \equiv \left(\frac{\partial \mathcal{L}_{c_{\psi}}}{\partial \dot{c}_{\psi}}\right)_{\rm R}
= i \overline{c}_{\psi} \gamma^0 = i c_{\psi}^{\dagger}.
\label{pi-psi-c}
\end{eqnarray}

By solving the Dirac equations $(i \gamma^{\mu} \partial_{\mu} - m)\psi = 0$ 
and $(i \gamma^{\mu} \partial_{\mu} - m)c_{\psi} = 0$, 
we obtain the solutions,
\begin{eqnarray}
&~& \psi(x) = \int \frac{d^3k}{\sqrt{(2\pi)^3 2k_0}}\sum_{s}
\left(a(\bm{k}, s) u(\bm{k}, s) e^{-i k x} + b^{\dagger}(\bm{k}, s) v(\bm{k}, s)e^{i k x}\right),
\label{psi-sol}\\
&~& \pi_{\psi}(x) = i \int \frac{d^3k}{\sqrt{(2\pi)^3 2k_0}}\sum_{s}
\left(a^{\dagger}(\bm{k}, s) u^{\dagger}(\bm{k}, s) e^{i k x} + b(\bm{k}, s) v^{\dagger}(\bm{k}, s) e^{-i k x}\right),
\label{pi-psi-sol}\\
&~& c_{\psi}(x) = \int \frac{d^3k}{\sqrt{(2\pi)^3 2k_0}}\sum_{s}
\left(c(\bm{k}, s) u(\bm{k}, s) e^{-i k x} + d^{\dagger}(\bm{k}, s) v(\bm{k}, s)e^{i k x}\right),
\label{cpsi-sol}\\
&~& \pi_{c_{\psi}}(x) = i \int \frac{d^3k}{\sqrt{(2\pi)^3 2k_0}}\sum_{s}
\left(c^{\dagger}(\bm{k}, s) u^{\dagger}(\bm{k}, s) e^{i k x} + d(\bm{k}, s) v^{\dagger}(\bm{k}, s) e^{-i k x}\right),
\label{pi-cpsi-sol}
\end{eqnarray}
where $s$ represents the spin state,
and $u(\bm{k}, s)$ and $v(\bm{k}, s)$ are Dirac spinors on the momentum space.
They satisfy the relations,
\begin{eqnarray}
\sum_s u(\bm{k}, s) \overline{u}(\bm{k}, s) = \Slashk + m,~~
\sum_s v(\bm{k}, s) \overline{v}(\bm{k}, s) = \Slashk - m,
\label{uv}
\end{eqnarray}
where $\overline{u}(\bm{k}, s) \equiv {u}^{\dagger}(\bm{k}, s)\gamma^0$,
$\overline{v}(\bm{k}, s) \equiv {v}^{\dagger}(\bm{k}, s)\gamma^0$
and $\Slashk = \gamma^{\mu} k_{\mu}$.

Using (\ref{pi-psi-c}), the Hamiltonian density is obtained as
\begin{eqnarray}
\mathcal{H}_{\psi, c_{\psi}} = \pi_{\psi} \dot{\psi} + \pi_{c_{\psi}} \dot{c}_{\psi} - \mathcal{L}_{\psi, c_{\psi}}
= - i \sum_{i=1}^{3} \overline{\psi} \gamma^i \partial_i \psi + m  \overline{\psi} \psi
- i \sum_{i=1}^{3} \overline{c}_{\psi} \gamma^i \partial_i c_{\psi} + m  \overline{c}_{\psi}c_{\psi}.
\label{H-psi-c}
\end{eqnarray}

The system is quantized by regarding variables as operators
and imposing the following relations 
on the canonical pairs $(\psi, \pi_{\psi})$ and $(c_{\psi}, \pi_{c_{\psi}})$,
\begin{eqnarray}
\{\psi^{\alpha}(\bm{x}, t), \pi_{\psi}^{\beta}(\bm{y}, t)\}
 = i \delta^{\alpha \beta}\delta^3(\bm{x}-\bm{y}),~~ 
[c_{\psi}^{\alpha}(\bm{x}, t), \pi_{c_{\psi}}^{\beta}(\bm{y}, t)]
 = i \delta^{\alpha \beta}\delta^3(\bm{x}-\bm{y}), 
\label{CCR-psi-c}
\end{eqnarray}
and others are zero.
Here, $\alpha$ and $\beta$ are spinor indices.
Or equivalently, the following relations are imposed on,
\begin{eqnarray}
&~& \{a(\bm{k}, s), a^{\dagger}(\bm{l}, s')\} = \delta_{s s'} \delta^3(\bm{k}-\bm{l}),~~ 
\{b(\bm{k}, s), b^{\dagger}(\bm{l}, s')\} = \delta_{s s'} \delta^3(\bm{k}-\bm{l}),
\label{CCR-ab-psi}\\
&~& [c(\bm{k}, s), c^{\dagger}(\bm{l}, s')] = \delta_{s s'} \delta^3(\bm{k}-\bm{l}),~~ 
[d(\bm{k}, s), d^{\dagger}(\bm{l}, s')] = - \delta_{s s'} \delta^3(\bm{k}-\bm{l}),
\label{CCR-cd-cpsi}
\end{eqnarray}
and others are zero.

By inserting (\ref{psi-sol}) -- (\ref{pi-cpsi-sol}) into (\ref{H-psi-c}), 
the Hamiltonian $H_{\psi, c_{\psi}}$ is written by
\begin{eqnarray}
&~& H_{\psi, c_{\psi}} = \int \mathcal{H}_{\psi, c_{\psi}} d^3x
= \int d^3k \sum_{s} k_0 
\left(a^{\dagger}(\bm{k}, s) a(\bm{k}, s) + b^{\dagger}(\bm{k}, s) b(\bm{k}, s)\right.
\nonumber \\
&~& ~~~~~~~~~~~~~~~~~~~~~~~~~~~~~~~~~~~~~~~~~~~~
\left. + c^{\dagger}(\bm{k}, s) c(\bm{k}, s) - d^{\dagger}(\bm{k}, s) d(\bm{k}, s)\right),
\label{H-abcd-psi-c}
\end{eqnarray}
where the sum of the zero point energies vanishes due to the cancellation between
contributions from ($\psi$, $\psi^{\dagger}$) and ($c_{\psi}$, $c_{\psi}^{\dagger}$).

The eigenstates for $H_{\psi, c_{\psi}}$ are constructed by acting
the creation operators $a^{\dagger}(\bm{k}, s)$, $b^{\dagger}(\bm{k}, s)$,
$c^{\dagger}(\bm{k}, s)$ and $d^{\dagger}(\bm{k}, s)$ on the vacuum state $| 0 \rangle$,
where $| 0 \rangle$ is defined by the conditions $a(\bm{k}, s)| 0 \rangle = 0$,
$b(\bm{k}, s)| 0 \rangle = 0$, $c(\bm{k}, s)| 0 \rangle = 0$ and $d(\bm{k}, s)| 0 \rangle = 0$.
The energy is positive semi-definite,
because the effect on the negative sign 
in front of $d^{\dagger}(\bm{k}, s) d(\bm{k}, s)$ in ${H}_{\varphi, c_{\varphi}}$
changes into an opposite one by the negative sign in the relation 
$[d(\bm{k}, s), d^{\dagger}(\bm{l}, s)] = -\delta_{s s'} \delta^3(\bm{k}-\bm{l})$.

We find that two fields separated by a space-like interval 
anti-commute or commute with each other as seen
from $\varDelta(x-y) = 0$ for $(x-y)^2 < 0$ and the relations,
\begin{eqnarray}
&~& \{\psi^{\alpha}(x), \overline{\psi}^{\beta}(y)\} 
= [c_{\psi}^{\alpha}(x), \overline{c}_{\psi}^{\beta}(y)] 
= \left(i \gamma^{\mu} \partial_{\mu} + m\right)^{\alpha\beta}
 \int \frac{d^3k}{(2\pi)^3 2k_0} \left(e^{-ik(x-y)} - e^{ik(x-y)}\right)
\nonumber \\
&~& ~~~~~~~~~~~ 
=\left(i \gamma^{\mu} \partial_{\mu} + m\right)^{\alpha\beta} i \varDelta(x-y)
\equiv i S^{\alpha\beta}(x-y),~~ 
\label{4D-CCR-psi-c1}\\
&~& \{\psi^{\alpha}(x), \psi^{\beta}(y)\} = 0,~~
\{\overline{\psi}^{\alpha}(x), \overline{\psi}^{\beta}(y)\} = 0,~~
[c_{\psi}^{\alpha}(x), c_{\psi}^{\beta}(y)] = 0,~~
[\overline{c}_{\psi}^{\alpha}(x), \overline{c}_{\psi}^{\beta}(y)] = 0,
\label{4D-CCR-psi-c2}\\
&~& [\psi^{\alpha}(x), c_{\psi}^{\beta}(y)] = 0,~~
[\psi^{\alpha}(x), \overline{c}_{\psi}^{\beta}(y)] = 0,~~
[\overline{\psi}^{\alpha}(x), c_{\psi}^{\beta}(y)] = 0,~~
[\overline{\psi}^{\alpha}(x), \overline{c}_{\psi}^{\beta}(y)] = 0.
\label{4D-CCR-psi-c3}
\end{eqnarray}
Hence, the microscopic causality also holds on.

The system contains negative norm states as seen from the relation
$[d(\bm{k}, s), d^{\dagger}(\bm{l}, s')]
= - \delta_{s s'} \delta^3(\bm{k}-\bm{l})$.
It is also shown that the system has fermionic symmetries 
and they can guarantee the unitarity of the system.

The $\mathcal{L}_{\psi, c_{\psi}}$ is invariant under the fermionic transformations,
\begin{eqnarray}
&~& \delta_{\rm F} \psi = r \zeta c_{\psi},~~
\delta_{\rm F} \psi^{\dagger} = 0,~~ 
\delta_{\rm F} c_{\psi}  = 0,~~
\delta_{\rm F} c_{\psi}^{\dagger} = r \zeta \psi^{\dagger},
\label{delta-F-psi-c}\\
&~& \delta_{\rm F}^{\dagger} \psi = 0,~~
\delta_{\rm F}^{\dagger} \psi^{\dagger} = r \zeta^{\dagger} c_{\psi}^{\dagger},~~
\delta_{\rm F}^{\dagger} c_{\psi} = - r \zeta^{\dagger} \psi,~~
\delta_{\rm F}^{\dagger} c_{\psi}^{\dagger} = 0
\label{delta-Fdagger-psi-c}
\end{eqnarray}
and the $U(1)$ transformation,
\begin{eqnarray}
\delta \psi = - iq\epsilon \psi,~~
\delta \psi^{\dagger} = i q \epsilon \psi^{\dagger},~~
\delta c_{\psi} = - iq\epsilon c_{\psi},~~
\delta c_{\psi}^{\dagger} = i q \epsilon c_{\psi}^{\dagger},
\label{U(1)-psi-c}
\end{eqnarray}
where $r = q^{1/2}$ and $q$ is the $U(1)$ charge of $\psi$ and $c_{\psi}$.
The corresponding generators are given by
\begin{eqnarray}
&~& Q_{\rm F} = - i \int d^3k \sum_{s} r 
\left(a^{\dagger}(\bm{k}, s) c(\bm{k}, s) - d^{\dagger}(\bm{k}, s) b(\bm{k}, s)\right),
\label{QF-psi-c}\\
&~& Q_{\rm F}^{\dagger} = - i \int d^3k \sum_{s} r 
\left(c^{\dagger}(\bm{k}, s) a(\bm{k}, s) - b^{\dagger}(\bm{k}, s) d(\bm{k}, s)\right),
\label{QF-dagger-psi-c}\\
&~& N_{\rm D} 
= \int d^3k \sum_s q
\left(a^{\dagger}(\bm{k}, s) a(\bm{k}, s) - b^{\dagger}(\bm{k}, s) b(\bm{k}, s) 
\right.
\nonumber \\
&~& ~~~~~~~~~~~~~~~~~~~~~~~~~~~~~~~
\left. + c^{\dagger}(\bm{k}, s) c(\bm{k}, s) 
+ d^{\dagger}(\bm{k}, s) d(\bm{k}, s)\right).
\label{ND}
\end{eqnarray}
We have the algebraic relations ${Q_{\rm F}}^2 = 0$, ${Q_{\rm F}^{\dagger}}^2 = 0$
and $\{Q_{\rm F}, Q_{\rm F}^{\dagger}\} = N_{\rm D}$.
We find that the $U(1)$ charge of particle corresponding $b^{\dagger}(\bm{k}, s) | 0 \rangle$
is opposite to that corresponding $a^{\dagger}(\bm{k}, s) | 0 \rangle$.
Hence, $a(\bm{k}, s)$ and $b^{\dagger}(\bm{k}, s)$ are 
regarded as the annihilation operator of particle and the creation operator of antiparticle, respectively.
In the same way, $c(\bm{k}, s)$ and $d^{\dagger}(\bm{k}, s)$ are 
regarded as the annihilation operator of bosonic particle 
and the creation operator of bosonic antiparticle, respectively.

It is easily understood that $\mathcal{L}_{\psi, c_{\psi}}$ is invariant under the transformations
(\ref{delta-F-psi-c}) and (\ref{delta-Fdagger-psi-c}), 
from the nilpotency of ${\bm \delta}_{\rm F}$ and ${\bm \delta}_{\rm F}^{\dagger}$ 
and the relations,
\begin{eqnarray}
\mathcal{L}_{\psi, c_{\psi}} 
= {\bm \delta}_{\rm F} {\bm \delta}_{\rm F}^{\dagger} \left(\mathcal{L}_{\psi}/q\right)
= - {\bm \delta}_{\rm F}^{\dagger} {\bm \delta}_{\rm F} \left(\mathcal{L}_{\psi}/q\right),
\label{delta-rel-psi}
\end{eqnarray}
where $\mathcal{L}_{\psi}$ is given by
\begin{eqnarray}
\mathcal{L}_{\psi} =  i \overline{\psi} \gamma^{\mu} \partial_{\mu} \psi -m \overline{\psi}\psi.
\label{L-psi}
\end{eqnarray}
The Hamiltonian density $\mathcal{H}_{\psi, c_{\psi}}$ is written in the $Q_{\rm F}$ and $Q_{\rm F}^{\dagger}$
exact forms such that 
\begin{eqnarray}
\mathcal{H}_{\psi, c_{\psi}} 
= \left\{Q_{\rm F}, \left\{Q_{\rm F}^{\dagger}, 
\mathcal{H}_{\psi}/q\right\}\right\}
= - \left\{Q_{\rm F}^{\dagger}, \left\{Q_{\rm F}, 
\mathcal{H}_{\psi}/q\right\}\right\},
\label{H-psi-c-exact}
\end{eqnarray}
where $\mathcal{H}_{\psi}$ is given by
\begin{eqnarray}
\mathcal{H}_{\psi} = - i \sum_{i=1}^{3} \overline{\psi} \gamma^i \partial_i \psi + m  \overline{\psi} \psi.
\label{H-psi}
\end{eqnarray}

To formulate our model in a consistent manner,
we impose the subsidiary conditions,
\begin{eqnarray}
Q_{\rm F} |{\rm phys}\rangle = 0,~~
Q_{\rm F}^{\dagger} |{\rm phys}\rangle = 0,~~
N_{\rm D} |{\rm phys}\rangle = 0,
\label{Phys-again2}
\end{eqnarray}
and find that all states, except for the vacuum state $| 0 \rangle$,
are unphysical through the quartet mechanism, in the similar way as the scalar fields in the previous section.

\section{Conclusions and discussions}

We have studied the quantization of systems that contain 
both ordinary fields with a positive norm
and their counterparts obeying different statistics,
and found that the systems have new type of fermionic symmetries 
and the unitarity of systems holds by imposing
subsidiary conditions on states.

The systems considered are unrealistic,
because they are empty leaving the vacuum state alone
as the physical state.
$Q_{\rm F}$ singlet fields are needed to realize our world.
For a system that $Q_{\rm F}$-singlets
and $Q_{\rm F}$-doublets coexist with exact fermionic symmetries,
the Lagrangian density is, in general, written in the form as
$\mathcal{L}_{\rm Total} = \mathcal{L}_{\rm S} + \mathcal{L}_{\rm D} + \mathcal{L}_{\rm mix}
= \mathcal{L}_{\rm S} + \bm{\delta}_{\rm F} \bm{\delta}_{\rm F}^{\dagger} (\Delta \mathcal{L})$.
Here, $\mathcal{L}_{\rm S}$, $\mathcal{L}_{\rm D}$ 
and $\mathcal{L}_{\rm mix}$
stand for the Lagrangian density for $Q_{\rm F}$-singlets, 
$Q_{\rm F}$-doublets
and interactions between $Q_{\rm F}$-singlets and $Q_{\rm F}$-doublets.
Under the subsidiary conditions 
$Q_{\rm F} |{\rm phys}\rangle =0$,
$Q_{\rm F}^{\dagger} |{\rm phys}\rangle =0$
and $N_{\rm D} |{\rm phys}\rangle =0$
on states,
all $Q_{\rm F}$-doublets become unphysical.
This system seems to be same as that described by 
$\mathcal{L}_{\rm S}$ alone,
because $Q_{\rm F}$-doublets do not give any dynamical effects
on $Q_{\rm F}$-singlets.
From this, we suppose that it is not possible to show 
the existence of $Q_{\rm F}$-doublets.
However, in a very special case, an indirect proof would be possible
through fingerprints left by symmetries in a fundamental theory.
The fingerprints are specific relations among parameters
such as a unification of coupling constants,
reflecting on underlying symmetries~\cite{YK}. 
This subject will be reexamined in the separate publication~\cite{YK2}.

\section*{Acknowledgments}
The author thanks Prof. T. Kugo for valuable discussions 
and useful comments, in particular,
the clarification of the difference between our system
with $OSp(2|2)$ symmetry and a system with the BRST symmetry.
This work was supported in part by scientific grants from the Ministry of Education, Culture,
Sports, Science and Technology under Grant No.~22540272.

\appendix

\section{$OSp(2|2)$ and $OSp(1,1|2)$}

The $OSp(2|2)$ is the group whose elements are generators
of transformations which leave the inner product
$x^2 + y^2 + 2 i \theta_1 \theta_2$.
Here, $x$ and $y$ are real numbers, 
and $\theta_1$ and $\theta_2$ are hermitian Grassmann numbers,
\begin{eqnarray}
\theta_1^{\dagger} = \theta_1,~~\theta_2^{\dagger} = \theta_,~~
{\theta_1}^2 = 0,~~{\theta_2}^2 = 0.
\label{theta12}
\end{eqnarray} 
The infinitesimal transformations are classified into following types.\\
(a) Rotation relating $x$ and $y$:
\begin{eqnarray}
\delta_{\rm r} x = - \epsilon_{\rm r} y,~~
\delta_{\rm r} y = \epsilon_{\rm r} x,~~
\delta_{\rm r} \theta_1 = 0,~~ \delta_{\rm r} \theta_2 = 0,
\label{R}
\end{eqnarray} 
where $\epsilon_{\rm r}$ is an infinitesimal real parameter.\\
(b) Rotation relating $\theta_1$ and $\theta_2$:
\begin{eqnarray}
\delta_{\rm r'} x = 0,~~
\delta_{\rm r'} y = 0,~~
\delta_{\rm r'} \theta_1 = - \epsilon_{\rm r'} \theta_2,~~
\delta_{\rm r'} \theta_2 = \epsilon_{\rm r'} \theta_1,
\label{R-theta}
\end{eqnarray} 
where $\epsilon_{\rm r'}$ is an infinitesimal real parameter.\\
(c) Fermionic transformations:
\begin{eqnarray}
&~& \delta_{1} x = - i \zeta_{1} \theta_2,~~
\delta_{1} y = i \zeta_{1} \theta_1,~~
\delta_{1} \theta_1 = \zeta_{1} x,~~ 
\delta_{1} \theta_2 = \zeta_{1} y,
\label{F1}\\
&~& \delta_{2} x = - i \zeta_{2} \theta_1,~~
\delta_{2} y = -i \zeta_{2} \theta_2,~~
\delta_{2} \theta_1 = \zeta_{2} y,~~ 
\delta_{2} \theta_2 = -\zeta_{2} x,
\label{F2}
\end{eqnarray} 
where $\zeta_1$ and $\zeta_2$ are Grassmann numbers.

By introducing four hermitian scalar fields,
we can construct a Lagrangian density 
with $OSp(2|2)$ invariance as follows,
\begin{eqnarray}
\mathcal{L}_{OSp(2|2)}
= \frac{1}{2} \left(\partial_{\mu} \phi_1 \partial^{\mu} \phi_1
+ \partial_{\mu} \phi_2 \partial^{\mu} \phi_2\right)
- \frac{1}{2} m^2 \left({\phi_1}^2 + {\phi_2}^2\right) 
+ i \partial_{\mu} c_1 \partial^{\mu} c_2
- i m^2 c_1 c_2,
\label{L-OSp2-2}
\end{eqnarray} 
where $\phi_1$ and $\phi_2$ are ordinary hermitian scalar fields
and $c_1$ and $c_2$ are fermionic hermitian scalar fields.

Using complex scalar fields defined by
\begin{eqnarray}
\varphi \equiv \frac{1}{\sqrt{2}} \left(\phi_1 + i \phi_2\right),~~
c_{\varphi} \equiv \frac{1}{\sqrt{2}} \left(c_1 + i c_2\right),
\label{varphi-c}
\end{eqnarray} 
the above Lagrangian density (\ref{L-OSp2-2}) is rewritten as
\begin{eqnarray}
\mathcal{L}_{OSp(2|2)}
= \partial_{\mu} \varphi^{\dagger} \partial^{\mu} \varphi
- m^2 \varphi^{\dagger} \varphi
+ \partial_{\mu} c_{\varphi}^{\dagger} \partial^{\mu} c_{\varphi}
- m^2 c_{\varphi}^{\dagger} c_{\varphi}.
\label{L-OSp2-2-varphi}
\end{eqnarray} 
The Lagrangian density (\ref{L-OSp2-2-varphi})
is just given by (\ref{L-varphi-c}).

For a reference sake,
we compare the above-mensioned system 
with a system of scalar fields with $OSp(1,1|2)$.
The $OSp(1,1|2)$ is the group whose elements are generators
of transformations which leave the inner product
$x^2 - y^2 + 2 i \theta_1 \theta_2$.
Notice that a negative sign exists in front of $y^2$.
The infinitesimal transformations are classified into following types.\\
(a) Boost relating $x$ and $y$:
\begin{eqnarray}
\delta_{\rm b} x = -\epsilon_{\rm b} y,~~
\delta_{\rm b} y = -\epsilon_{\rm b} x,~~
\delta_{\rm b} \theta_1 = 0,~~ \delta_{\rm b} \theta_2 = 0,
\label{B}
\end{eqnarray} 
where $\epsilon_{\rm b}$ is an infinitesimal real parameter.\\
(b) Rotation relating $\theta_1$ and $\theta_2$:
\begin{eqnarray}
\delta_{\rm r'} x = 0,~~
\delta_{\rm r'} y = 0,~~
\delta_{\rm r'} \theta_1 = - \epsilon_{\rm r'} \theta_2,~~
\delta_{\rm r'} \theta_2 = \epsilon_{\rm r'} \theta_1,
\label{R-theta-again}
\end{eqnarray} 
where $\epsilon_{\rm r'}$ is an infinitesimal real parameter.\\
(c) Fermionic transformations:
\begin{eqnarray}
&~& \delta_{\rm B} x = \lambda \theta_1,~~
\delta_{\rm B} y = - \lambda \theta_1,~~
\delta_{\rm B} \theta_1 = 0,~~ 
\delta_{\rm B} \theta_2 = i \lambda (x+y),
\label{B1}\\
&~& \overline{\delta}_{\rm B} x = \lambda \theta_2,~~
\overline{\delta}_{\rm B} y = - \lambda \theta_2,~~
\overline{\delta}_{\rm B} \theta_1 = - i \lambda (x+y),~~ 
\overline{\delta}_{\rm B} \theta_2 = 0,
\label{B2}
\end{eqnarray} 
where $\lambda$ is a Grassmann numbers with $\lambda^{*} = - \lambda$.

By introducing four hermitian scalar fields,
we can construct a Lagrangian density 
with $OSp(1,1|2)$ invariance as follows,
\begin{eqnarray}
\mathcal{L}_{OSp(1,1|2)}
= \frac{1}{2} \left(\partial_{\mu} \phi_3 \partial^{\mu} \phi_3
- \partial_{\mu} \phi_0 \partial^{\mu} \phi_0\right)
- \frac{1}{2} m^2 \left({\phi_3}^2 - {\phi_0}^2\right) 
+ i \partial_{\mu} c_1 \partial^{\mu} c_2
- i m^2 c_1 c_2,
\label{L-OSp1-1-2}
\end{eqnarray}
where $\phi_3$ is an ordinary hermitian scalar field,
$\phi_0$ is a hermitian scalar field with a negative norm,
and $c_1$ and $c_2$ are fermionic hermitian scalar fields.

Using hermitian scalar fields defined by
\begin{eqnarray}
B \equiv \frac{1}{\sqrt{2}} \left(\phi_3 + \phi_0\right),~~
\phi \equiv \frac{1}{\sqrt{2}} \left(\phi_3 - \phi_0\right),
\label{B-phi}
\end{eqnarray} 
the above Lagrangian density (\ref{L-OSp1-1-2}) is rewritten as
\begin{eqnarray}
\mathcal{L}_{OSp(1,1|2)}
= \partial_{\mu} B \partial^{\mu} \phi
- m^2 B \phi
+ i \partial_{\mu} \overline{c} \partial^{\mu} c
- i m^2 \overline{c} c,
\label{L-OSp1-1-2-B}
\end{eqnarray} 
where $\overline{c} = c_1$ and $c = c_2$.
The interacting model containing $\mathcal{L}_{OSp(1,1|2)}$
as a free part has been constructed and studied~\cite{F2,F3}.

The Lagrangian density (\ref{L-OSp1-1-2-B}) is invariant
under the following fermionic transformations,
\begin{eqnarray}
&~& \delta_{\rm B} \phi = \lambda c,~~
\delta_{\rm B} c = 0,~~
\delta_{\rm B} \overline{c} = i \lambda B,~~ 
\delta_{\rm B} B = 0,
\label{BRST}\\
&~& \overline{\delta}_{\rm B} \phi = \lambda \overline{c},~~
\overline{\delta}_{\rm B} c= - i \lambda B~~
\overline{\delta}_{\rm B} \overline{c} = 0,~~ 
\overline{\delta}_{\rm B} B = 0.
\label{anti-BRST}
\end{eqnarray} 
They correspond to the BRST and anti-BRST transformations, respectively.
The following algebraic relations hold:
\begin{eqnarray}
{Q_{\rm B}}^2 = 0,~~ {\overline{Q}_{\rm B}}^2 = 0,~~
\{Q_{\rm B}, \overline{Q}_{\rm B}\} = 0,
\label{BRST-algebra}
\end{eqnarray}
where $Q_{\rm B}$ and $\overline{Q}_{\rm B}$ are 
the BRST and the anti-BRST charges given by
\begin{eqnarray}
\delta_{\rm B} \Phi = i[\lambda Q_{\rm B}, \Phi],~~
\overline{\delta}_{\rm B} \Phi = i[\lambda \overline{Q}_{\rm B}, \Phi].
\label{BRST-charge}
\end{eqnarray}

The Lagrangian density (\ref{L-OSp1-1-2-B}) is rewritten by
\begin{eqnarray}
\mathcal{L}_{OSp(1,1|2)}
= \bm{\delta}_{\rm B} 
\left(- i \partial_{\mu} \overline{c} \partial^{\mu} \phi
+ i m^2 \overline{c} \phi\right)
= \bm{\delta}_{\rm B} \overline{\bm{\delta}}_{\rm B}
\left(- \frac{i}{2}\partial_{\mu} \phi \partial^{\mu} \phi
+ \frac{i}{2} m^2 \phi \phi\right),
\label{L-OSp1-1-2-delta}
\end{eqnarray} 
where $\bm{\delta}_{\rm B}$ and $\overline{\bm{\delta}}_{\rm B}$ 
are defined by 
$\delta_{\rm B} = \lambda \bm{\delta}_{\rm B}$
and $\overline{\delta}_{\rm B} 
= \lambda \overline{\bm{\delta}}_{\rm B}$, respectively.

Finally, we point out that the Lagrangian density (\ref{L-OSp1-1-2-delta})
consists of the gauge-fixing term and the Faddeev-Popov ghost term
for the system of $\phi$ with an empty dynamics.
The system with the empty action integral $S=0$ has the
invariance under the local transformation
$\phi(x) \to \phi_{\Lambda} = \phi(x) + \Lambda(x)$, 
and after taking the gauge-fixing condition,
\begin{eqnarray}
f(\phi_{\Lambda}(x)) 
= (\partial_{\mu} \partial^{\mu} + m^2) \phi(x) = 0,
\label{gf}
\end{eqnarray} 
we obtain the Lgrangian density,
\begin{eqnarray}
\mathcal{L}_{\rm gf+gh}
= \bm{\delta}_{\rm B} 
\left(- i \overline{c} (\partial_{\mu} \partial^{\mu} + m^2)\phi\right)
= - B(\partial_{\mu} \partial^{\mu} + m^2)\phi
-  i \overline{c} (\partial_{\mu} \partial^{\mu} + m^2)c.
\label{L-gf}
\end{eqnarray} 
The Lagrangian density (\ref{L-gf}) becomes $\mathcal{L}_{OSp(1,1|2)}$
after the partial integration in the action integral.

\end{document}